\documentclass[useAMS,usenatbib]{mn2e}
\usepackage{amsmath}
\usepackage{amssymb}
\usepackage{multirow}
\usepackage{graphicx}
\usepackage{epstopdf}

\def\simlt{\lower.5ex\hbox{$\; \buildrel < \over \sim \;$}}
\def\simgt{\lower.5ex\hbox{$\; \buildrel > \over \sim \;$}}

\newcommand{\bd}{\begin{displaymath}}
\newcommand{\ed}{\end{displaymath}}
\newcommand{\be}{\begin{equation}}
\newcommand{\ee}{\end{equation}}
\newcommand{\beqa}{\begin{eqnarray}}
\newcommand{\eeqa}{\end{eqnarray}}

\newcommand{\mpc}{\rm Mpc}

\newcommand{\vbc}{v_{\rm bc}}

\newcommand{\Mcool}{M_{\rm cool}}
\newcommand{\Mcrit}{M_{\rm crit}}
\newcommand{\Zcrit}{Z_{\rm crit}}

\title[21-cm BAO signature of enriched low-mass galaxies] {The 21-cm BAO
  signature of enriched low-mass galaxies during cosmic reionization}
\author[Cohen, Fialkov \& Barkana] {Aviad Cohen$^{1}$\thanks{E-mail:
    aviadc11@gmail.com}, Anastasia Fialkov$^{2,3}$,
  Rennan Barkana$^{1,4,5}$ \\
  $^{1}$ Raymond and Beverly Sackler School of Physics and Astronomy,
  Tel Aviv University, Tel Aviv 69978, Israel\\
  $^{2}$ Departement de Physique, Ecole Normale Sup\'{e}rieure, CNRS,
  24 rue Lhomond, Paris, 75005 France\\
  $^{3}$ Institute for Theory and Computation, Harvard University, 60 Garden Street, $MS-51$, Cambridge, MA, 02138 U.S.A.\\
  $^{4}$ Sorbonne Universit\'{e}s, Institut Lagrange de
  Paris (ILP), Institut d'Astrophysique de Paris, UPMC Univ Paris 06/CNRS,\\
  \hspace{.065in} 98 bis Boulevard Arago, 75014 Paris, France\\
  $^{5}$ Department of Astrophysics, University of Oxford, Denys Wilkinson Building, Keble Road, Oxford OX1 3RH, UK
  }%%%

\begin{document}
\pagerange{\pageref{firstpage}--\pageref{lastpage}} \pubyear{2015}
\maketitle

\label{firstpage}

\begin{abstract} 

  Studies of the formation of the first stars have established that
  they formed in small halos of $\sim 10^5 - 10^6 M_{\odot}$ via
  molecular hydrogen cooling. Since a low level of ultraviolet
  radiation from stars suffices to dissociate molecular hydrogen,
  under the usually-assumed scenario this primordial mode of star
  formation ended by redshift $z \sim 15$ and much more massive halos
  came to dominate star formation. However, metal enrichment from the
  first stars may have allowed the smaller halos to continue to form
  stars. In this Letter we explore the possible effect of star
  formation in metal-rich low-mass halos on the redshifted 21-cm
  signal of neutral hydrogen from $z = 6-40$. These halos are
  significantly affected by the supersonic streaming velocity, with
  its characteristic baryon acoustic oscillation (BAO) signature.
  Thus, enrichment of low-mass galaxies can produce a strong signature
  in the 21-cm power spectrum over a wide range of redshifts,
  especially if star formation in the small halos was more efficient
  than suggested by current simulations. We show that upcoming radio
  telescopes can easily distinguish among various possible scenarios.
  
\end{abstract}


\section{Introduction}

\label{Sec:Intro}

The first generation of stars is thought to have formed out of
pristine gas via radiative cooling of molecular hydrogen, H$_2$, in
halos with mass above $\sim10^5$ M$_{\odot}$
\citep[e.g.,][]{Tegmark:1997, Bromm:2002, Yoshida:2003}. Radiative
cooling of atomic hydrogen (and helium) could only happen in the rarer
halos that were heavier than $\sim 3 \times 10^7 M_{\odot}$,
corresponding to virial temperatures above $\sim 10^4$~K
\citep[e.g.,][]{Barkana:2001}.

Radiation emitted by the first stars likely had a dramatic impact on
subsequent star formation (SF). In particular, photons in the
Lyman-Werner (LW) band dissociated H$_2$ and suppressed SF
\citep{Haiman:1997}. Simulations have confirmed that, in the presence
of LW radiation, the minimum mass of a halo in which primordial gas
can radiatively cool increases \citep{Machacek:2001, Wise:2007,
  O'Shea:2008, Visbal:2014}.  This has motivated the usually-assumed
paradigm that SF in molecular-cooling halos was
significantly suppressed at $z \sim 20-25$ and completely cut off at
$z \sim 10-15$ \citep{Ahn:2009,Holzbauer:2011,Fialkov:2013}, with the
more massive atomic-cooling halos likely taking over SF
during the early stages of cosmic reionization.

SF in low-mass halos has major observational implications. In general,
the most promising probe of early galaxies is the redshifted 21-cm
line of neutral hydrogen \citep[e.g.,][]{Pritchard:2012}. The
brightness temperature of the 21-cm signal depends on SF through the
environmental effects of radiative backgrounds (Lyman-$\alpha$, X-ray,
and ionizing, in particular). Efforts to detect this signal are
underway in low-frequency radio observatories whose main focus at
present is to probe mid-to-late reionization ($z \sim 6-12$)
\citep[e.g.,][]{vanHaarlem:2013, Bowman:2013, Ali:2015}, while future
experiments are being designed to probe the signal from even higher
redshifts \citep{Mellema:2013}.

A particularly important effect on small halos is the supersonic
streaming velocity of baryons relative to the cold dark matter, often
referred to as $\vbc$ \citep{Tseliakhovich:2010}. The baryonic wind
suppresses gas infall and SF in $\sim 10^6$ M$_\odot$ halos
\citep[e.g.,][]{Tseliakhovich:2010, Dalal:2010, Tseliakhovich:2011,
  Maio:2011, Stacy:2011, Greif:2011, Fialkov:2012, mcquinn12, naoz2,
  Fialkov:2014c}. The power spectrum of relative velocity exhibits a
strong signature of baryon acoustic oscillations (BAOs) that were
imprinted in the velocity field at recombination.  Through the effect
of $\vbc$ on SF, the BAOs are inherited by the spatial distribution of
stars \citep{Dalal:2010, Visbal:2012}. This signature, potentially
detectable in 21-cm fluctuations, is expected to remain only as long
as small halos (well below the atomic-cooling threshold) contribute
significant SF.

Depending on their masses, the first generation of metal-free (Pop
III) stars ended their lives either via supernova explosions or by
direct collapse to a black hole \citep{Woosley:2015}. If a star died
in an explosion, metals produced in its core were flung out into its
parent galaxy and diffused into the surrounding intergalactic medium.
A single explosion was likely enough to raise the gas metallicity
within a significant volume above the critical level, $\Zcrit\sim
10^{-4}\, Z_{\odot}$, at which metal-line cooling becomes efficient
\citep{Karlsson:2013}. Since metals allow cooling down to temperatures
as low as H$_2$ cooling (and even lower), early metal enrichment
suggests a way to avoid the above-mentioned scenario in which small
halos become irrelevant at quite high redshifts. Unlike molecular
cooling in halos of similar masses, metal cooling is not strongly
affected by the LW radiation. Thus, even when the LW background builds
up, these halos can continue to form stars. The process that {\em
  does}\/ eventually sterilize these light halos is photoheating
feedback \citep[e.g.,][]{R86,WHK97,NS00,Sobacchi:2013}; once stars
emit ionizing radiation, the intergalactic gas that is photoheated
above $10^4$~K stops accreting onto halos below $\sim 10^8 - 10^9
M_\odot$, thus quenching subsequent SF within them.

While the details of Pop III stellar evolution and of metallicity
dispersal and clumping are complex and uncertain, recent simulations
\citep{Wise:2014,Jeon:2014} boost this novel scenario. These numerical
studies of early chemical feedback found that metal-enriched SF in
halos down to $M \sim 10^6 M_\odot$ can account for $\sim 30\%$ of the
ionizing photon budget of reionization. Motivated by the possibility
of avoiding LW feedback via metal cooling, in this Letter we explore
the 21-cm signatures of SF that continues in low-mass halos until
reionization.

\section{Model and Methods}
\label{Sec:Methods}

We use a semi-numerical simulation \citep{Visbal:2012} to follow the
non-linear evolution of the 21-cm signal in a $(768~\mpc)^3$ volume
with $(3~\mpc)^3$ cells. We start from a random realization of the
initial density and velocity fields, with standard cosmological
parameters \citep{Ade:2013}. A sub-grid model yields SF
and the resulting radiation from each cell. Integration over sources
yields the inhomogeneous backgrounds of X-ray, Ly$\alpha$ and ionizing
radiation, and the resulting 21-cm intensity.

In order to include the photoheating feedback on SF, we begin with the
simulation-based formula of \citet{Sobacchi:2013} for the minimum halo
mass $\Mcrit$ that allows significant gas accretion within an ionized
region:
\begin{equation}
  \Mcrit=M_0 J_{21}^a\left( \frac{1+z}{10}\right)^b\left[ 1-\left( 
      \frac{1+z}{1+z_{\rm IN}}\right)^c \right]^d\ ,
\label{Eq:Mcrit}
\end{equation}
a function of redshift $z$, the intensity J$_{21}$ of ionizing
radiation~\footnote{Noting that $a$ is small and J$_{21}$ depends only
  weakly on redshift (Fig.~7 of \citet{Sobacchi:2013}), we simply set
  J$_{21}=0.5$.} in units of $10^{-21}$ erg s$^{-1}$ Hz$^{-1}$
cm$^{-2}$ sr$^{-1}$, the redshift $z_{\rm IN}$ at which the region was
reionized, and the parameters $(M_0, a, b, c, d) = (2.8\times10^9
M_{\odot}, 0.17, -2.1, 2, 2.5)$~.

Eq.~(\ref{Eq:Mcrit}) considers the effect of feedback on a
fully-ionized volume only, while in our simulation some cells are
partially ionized by internal sources. We thus modify the prescription
of \citet{Sobacchi:2013} to more realistically include the feedback on
partially ionized cells. At any given redshift $z$, we set $z_{\rm
  IN}$ to be the redshift at which the ionization fraction in the cell
first passed half of its current value at $z$. This improved
prescription strengthens feedback and delays reionization by $\Delta
z\sim 1$. We note that in scenarios in which low-mass halos form stars
up until their region is reionized, the large gap between the
characteristic halo masses of metal cooling ($10^6 M_\odot$) and of
photoheating feedback ($M_0 = 2\times10^9 M_{\odot}$) results in
extreme, sudden photoheating feedback that must be computed precisely.

In our simulation a halo forms stars if it is heavy enough to both
radiativly cool its gas and resist photoheating feedback. Hence, the
minimum halo mass for SF is $M_{\rm min}= {\rm max}[\Mcool, \Mcrit]$.
A typical cell in our simulation begins with $z=z_{\rm IN}$ and
$\Mcrit=0$. Thus, $M_{\rm min}=\Mcool$, until reionization of the cell
proceeds enough that the difference between $z_{\rm IN}$ and $z$
increases until $\Mcrit$ exceeds $\Mcool$, leading to $M_{\rm
  min}=\Mcrit$.  $M_{\rm min}$ is used to determine the fraction of
gas in star-forming halos, $f_{\rm gas}$ \citep{Tseliakhovich:2011}.
For a partially ionized cell,
\begin{equation}
  f_{\rm gas}=f_{\rm gas,neut}(1-x_{\rm ion})+f_{\rm gas,ion}x_{\rm ion},
\end{equation}
where $x_{\rm ion}$ is the ionized fraction of the cell, $f_{\rm
  gas,neut}$ is the star-forming gas fraction in this cell assuming no
photoheating feedback and $f_{\rm gas,ion}$ is the same fraction with
the feedback. In addition, the effective $\Mcool$ increases in regions
with non-zero $\vbc$ \citep{Fialkov:2012}.
    
Metal cooling in low-mass early halos has not been numerically
investigated as systematically as the case of molecular cooling, plus
it is in any case far more uncertain due to the additional dependence
on the production and dispersal of metals. We conservatively assume
that metal cooling occurs in the same halos as H$_2$ cooling (i.e.,
down to a circular velocity of $4.2$ km s$^{-1}$), although metal
cooling can occur at even lower temperatures. One possible difference
between the two cooling channels is the mass dependence of the SF
efficiency $f_*(M)$ at the low-mass end. For H$_2$ cooling, we assume
a gradual low-mass cutoff as assumed by \citet{Fialkov:2013} based on
simulations \citep{Machacek:2001}:
\begin{equation}
 f_*(M) = \left\{ \begin{array}{ll}
         f_* & M_{\rm atomic} < M\ ,\\
         f_*\frac{\log(M/M_{\rm min})}{\log(M_{\rm atomic}/M_{\rm min})} & 
         M_{\rm min} < M < M_{\rm atomic}\ ,\\
        0 & \mbox{otherwise,}\end{array} \right.
        \label{eq:fstar1}
\end{equation}
where $f_*$ is the efficiency (assumed constant) above the minimum
halo mass for atomic cooling, $M_{\rm atomic}$.  However, this is not
expected to apply to metal cooling. Thus, we also consider a simple,
sharp cutoff at $M_{\rm min}$, with a constant SF efficiency $f_*$ at
all $M > M_{\rm min}$. We note that even in the H$_2$ cooling case,
the applicability of Eq.~(\ref{eq:fstar1}) is quite uncertain since it
is based on simulations that are far from demonstrating numerical and
physical convergence.  In addition, the value of $f_*$ is highly
unconstrained.  Some observations suggest a decreasing SF efficiency
in metal-poor galaxies \citep{Shi:2014}, which may be consistent with
low $f_*$ values in simulations
\citep{Jeon:2014,Wise:2014,O'Shea:2015} due to the overall effect of
supernovae and radiative feedbacks. We adopt $f_*\sim5\%$ as a typical
efficiency.

We wish to compare to a model that is based closely on current
simulations. \citet{O'Shea:2015} presented a fitting function for the
fraction of halos containing metal-enriched stars at the final
redshift of their simulations, $f_{\rm occ}(M)$ (the halo occupancy
fraction).
%\begin{equation}
%f_{\rm occ}(M) =\left[ 1+\left( 2^{\alpha/3}-1\right)\left(\frac{M}{M_c} \right)^{-\alpha} \right]^{-3/\alpha}
%\label{eq:focc} 
%\end{equation}
%with $M_c=6\times10^7M_\odot$ and $\alpha=1.5$. 
This formula implies that halos below $\sim10^7M_\odot$ contain almost
no stars. Based on the results of \citet{Wise:2014} (see their
Figure~3), we adopt a smooth transition from high occupancy to the
$f_{\rm occ}(M)$ formula at redshifts for which the reionization
fraction goes from $\sim 1\%$ to $\sim 15\%$.  Based on the
simulations, in this case we also adopt a low SF efficiency of
$0.6\%$.

It is important to note that current simulations are limited by
resolution, given that $\sim 500$ particles per halo are needed in
order to determine overall halo properties (such as halo mass) to
within a factor of two \citep{Springel:2003}; for better accuracy or
to determine properties such as SF, more particles are required. In
\citet{Wise:2014} and \citet{O'Shea:2015}, the mass of a 500-particle
halo is $9\times10^5 M_\odot$ and $1.4\times10^7 M_\odot$,
respectively. Simulations with much higher dark matter resolution
\citep{Jeon:2015} find some (low-efficiency) SF in halos down to $10^6
M_\odot$ at relatively low redshifts, though such simulations can only
follow a tiny cosmic volume that is not representative. Some of our
models below assume significantly more SF than is suggested by current
simulations; our goal is to show that upcoming observations can easily
distinguish among these various possibilities.

%%%

%Thus, we consider one more possibility for the SF efficiency in halos below atomic cooling mass:
%\begin{equation}
%\label{eq:fstarSim} 
%f_*(M,z) = \frac{1}{2}f_*\left[ \left(1-f_{\rm occ}(M) \right)\tanh{\left( \frac{z-17}{3}\right) } + 1+f_{\rm occ}(M) \right] 
%\end{equation}

%%%

Another high-redshift process that is poorly constrained is heating.
The spectral energy distribution (SED) of high-z X-ray sources is
often modeled by a soft power-law spectrum \citep{Furlanetto:2006}.
However, the spectrum of X-ray binaries, the known source population
most likely to dominate high-redshift X-rays, is expected to be rather
hard \citep{Fragos:2013}. Illuminated by the hard (compared to soft)
X-rays, the Universe heats up more slowly and uniformly, with the
heating transition (defined as the time when the mean gas temperature
equals that of the cosmic microwave background) delayed due to the
larger mean free path of the hard X-ray photons \citep{Fialkov:2014}.

\section{Results}
\label{Sec:results}

We show predictions for the 21-cm power spectrum (PS) in a number of
scenarios, comparing various possibilities including significant metal
cooling in small halos. We consider large scales that are feasible to
observe and include the characteristic scale of $\sim 120$~Mpc of the
BAO signature from $\vbc$. We cover a wide range of redshifts and
cosmic events, exploring how the possibility of metal cooling
increases the range of possible 21-cm signals compared to previous
expectations.

In order to survey the parameter space and isolate various effects
(see section~\ref{Sec:Methods}), we compare six models/cases:
\begin{itemize}
\item Our ``Maximal'' case: hard X-ray SED, sharp $f_*(M)$ cutoff
  (i.e., constant $f_*$ down to $M_{\rm min}$), with photoheating
  feedback.  This exemplifies a case with effective metal-cooling but
  ineffective stellar feedback, allowing efficient SF in halos as
  small as the minimum halo mass for H$_2$ cooling.
  %%%
\item The ``Simulation-Based'' case: Like the Maximal, except with the
  smooth transition to $f_{\rm occ}(M)$ and a ten times lower SF
  efficiency.  This case represents current simulation results.  %%%
\item The ``Gradual'' case: Like the Maximal, except with a gradual
  cutoff in $f_*(M)$ (Eq.~\ref{eq:fstar1}). For the metal-cooling
  case, this illustrates some of the range of uncertainty in the
  effectiveness of SF in small halos; it also corresponds to the case
  of H$_2$ cooling in the limit of ineffective LW feedback.
\item The ``No-Feedback'' case: Like the Maximal, 
but without photoheating feedback. This is a comparison case, useful for 
isolating the effect of photoheating feedback.
\item The ``Soft'' case: Like the Maximal, except with a soft X-ray
  SED. This case illustrates the effect of one of the key uncertain
  parameters, the spectrum of early X-ray sources.
\item The ``Atomic'' case: Like the Maximal, but assuming star
  formation only in halos above the minimum mass for atomic cooling.
  This is the limiting case of ineffective SF in small halos due to LW
  or internal feedback; it provides a comparison case for highlighting
  the impact of small halos.
\end{itemize}
For each model the SF efficiency is normalized so that the escape
fraction of ionizing photons is $\sim 20\%$ and the total optical
depth to reionization is $\tau = 0.066$ \citep{Ade:2015} (except for
the Simulation-Based case, in which we set $f_*=0.006$ and assume the
resulting required ionizing efficiency). %%%
If with this choice of $f_*$ reionization ends too late, we set it to
yield a reionization fraction of at least 95\% by $z=6$, while making
sure that the optical depth is within the 2-$\sigma$ error bars of
{\it Planck} \citep{Ade:2015}.

Since the streaming velocity causes the most interesting observational
signature of small halos, we compare the predictions of each model
including or excluding the effect of $\vbc$. The effect of $\vbc$ is
expected to be stronger in models with weak feedback, where stars form
efficiently in small halos. This effect should thus decrease in the
following order: No-Feedback, Maximal/Soft, Gradual, Simulation-Based
and Atomic. %%%

\begin{figure*}
  \includegraphics[width=3.2in]{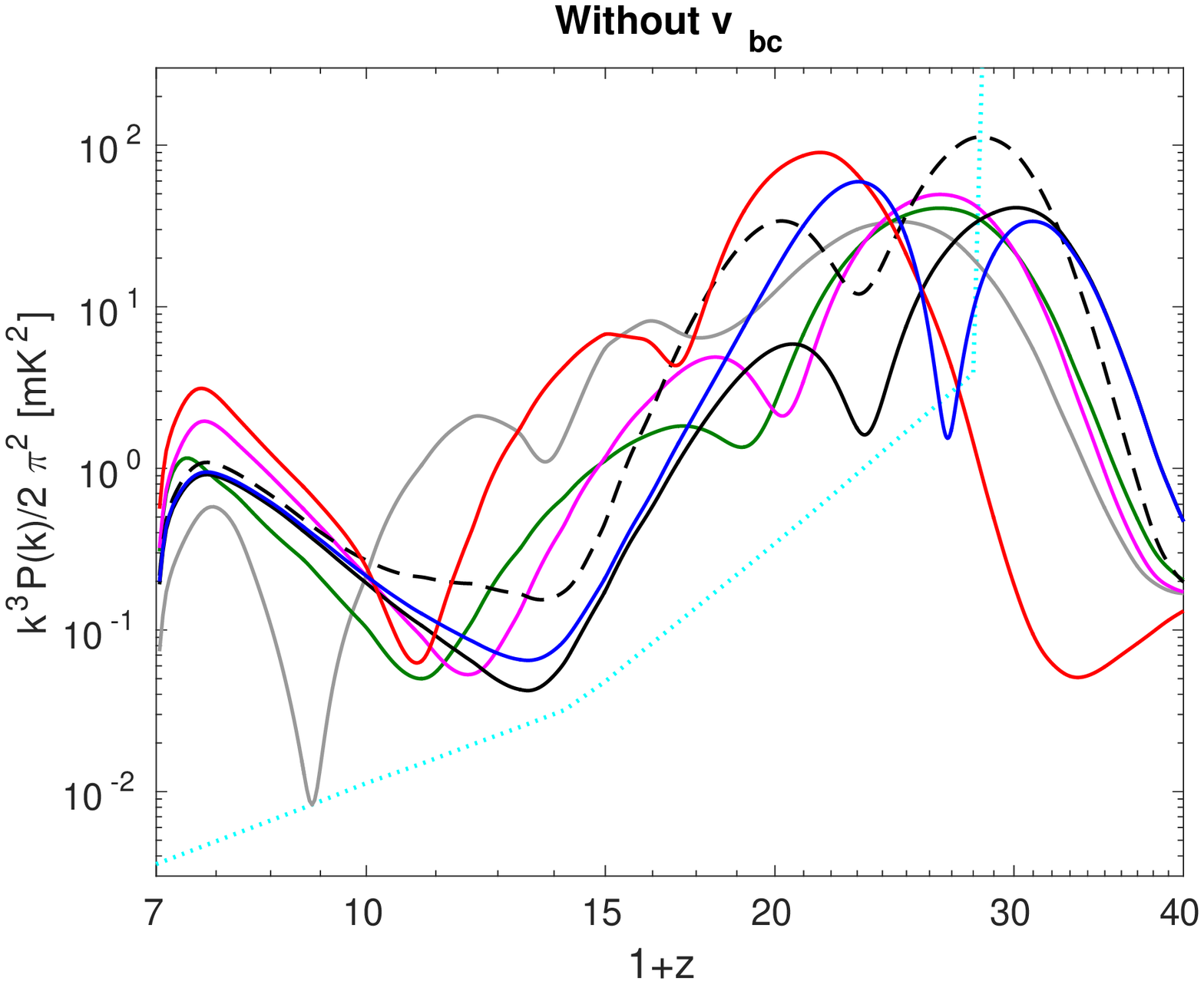}
  \includegraphics[width=3.2in]{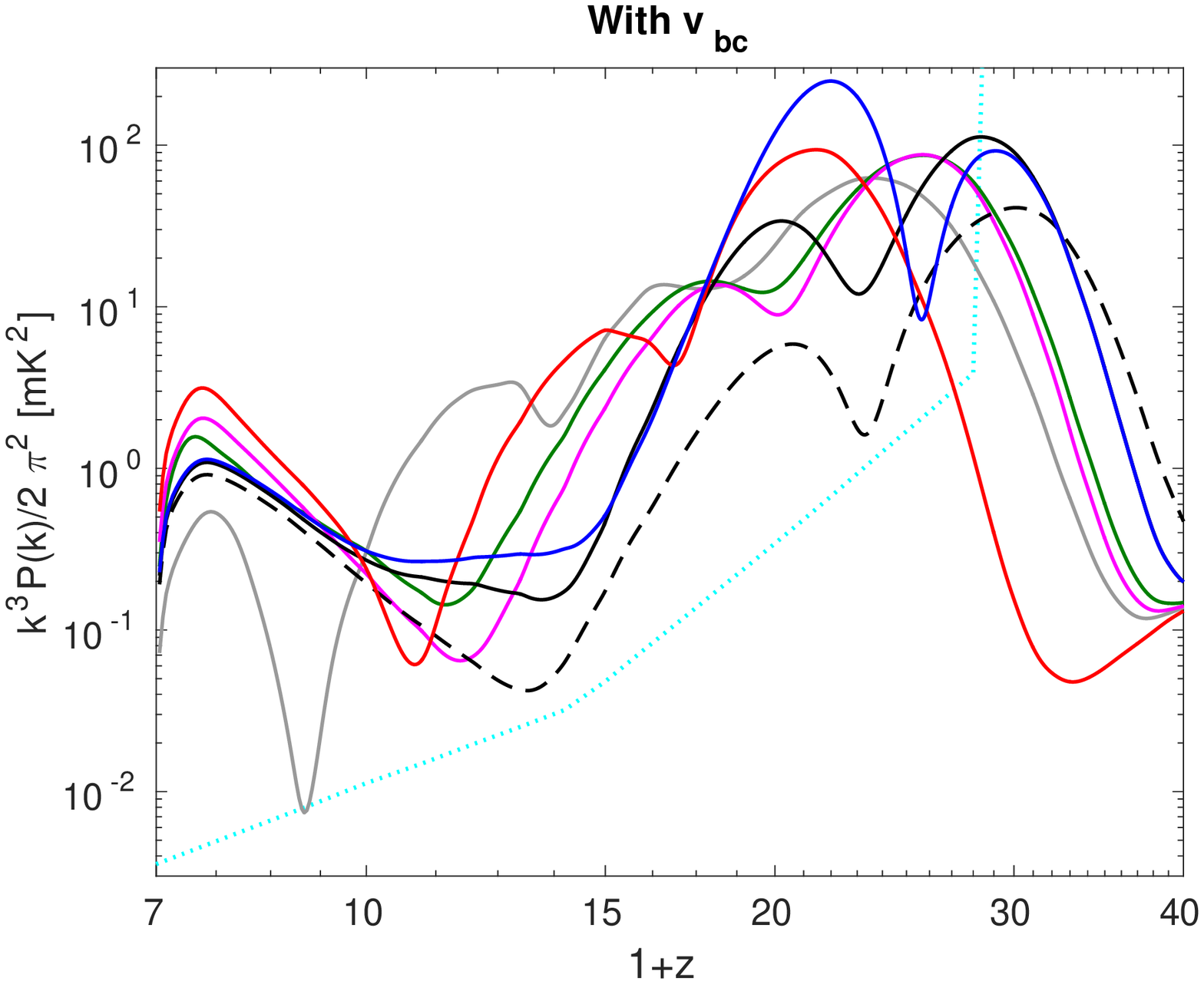}
  \caption{{\bf The (spherically-averaged) 21-cm power spectrum as a
      function of redshift.} We show the amplitude at $k=0.05
    ~\mpc^{-1}$ with (right panel) or without (left panel) $\vbc$. We
    consider our Maximal (black solid), Simulation-Based (gray),
    Gradual (magenta), No-Feedback (green), Soft (blue), or Atomic
    (red) models. For easier comparison, in each panel we also show
    the PS for the Maximal case from the other panel (black dashed).
    The cyan dotted line shows the power spectrum of the thermal noise
    for the SKA1 assuming a single beam, integration time of 1000
    hours, and 10 MHz bandwidth.}%%%
  \label{fig:1}
\end{figure*}

%Even focusing as we do here on the spherically-averaged 21-cm PS, we
%still have a 2D function (amplitude versus $k$ and $z$) for each
%model. 
We begin by showing the overall redshift evolution of large-scale
power (on the BAO scale) in Figure~\ref{fig:1}, for our six cases with
or without the streaming velocity. The comparison among the various
cases is made more complex by the fact that a number of distinct
sources of 21-cm fluctuations contribute at various times, and also
the models are normalized based on observational limits at mid-to-late
reionization (i.e., at the low-redshift end of the figure). A generic
plot of this type for the 21-cm PS has three peaks
\citep{BL05b,Jonathan07,Pritchard:2008}: the high-redshift peak, at $z
\sim 20-30$, is due to Ly$\alpha$ fluctuations; the mid-redshift peak,
at $z \sim 15-22$, is due to heating fluctuations; and the
low-redshift peak, at $z\sim7$ in these models, is due to ionization
fluctuations.

Without $\vbc$, models with more massive halos have stronger 21-cm
fluctuations, since such halos are more strongly biased, and show
lower-redshift peaks, since such halos form later. This can be seen by
comparing the Atomic, Gradual, and Maximal cases. The streaming
velocity has almost no effect on the Atomic case, but greatly enhances
the large-scale clustering of small halos, thus lifting the peaks in
the Gradual and Maximal cases to the same level as the Atomic case
(for the Ly$\alpha$ peak) or even higher (for the heating peak). We
note that the increased clustering (due to the spatial fluctuations in
$\vbc$) is added on top of the increased bias (due to the overall
$\vbc$-induced suppression of SF in low-mass halos). 

In the Simulation-Based case all features of the power spectrum are
shifted toward lower redshifts compared to the Maximal model due to
the slower build-up of stellar populations.  For this case there is an
extra peak at $z\sim14$ due to the transition to very low SF
efficiency in small halos (though this added feature is sensitive to
the assumed speed of this transition). In addition, fluctuations
during reionization are significantly suppressed since the much lower
SF efficiency results in colder reionization (for a fixed X-ray to SF
efficiency ratio). By comparing to the Maximal case, it can be seen
that there is an amplification of the fluctuations due to the $\vbc$
at high redshifts, while at redshifts lower than $\sim 14$ the $\vbc$
effect is negligible due to the transition to strong stellar feedback.
%%%

Comparing the Soft model to the Maximal, with or without $\vbc$,
shows that the main influence of the X-ray spectrum (which affects
only the heating fluctuations) is on the amplitude of the heating
peak; in the case of hard X-rays this peak is quite weak, and
disappears completely on smaller scales \citep{Fialkov:2014b}. The
No-Feedback model artificially keeps low-mass halos forming stars
until the end of reionization; the low-redshift normalization thus
forces this model to delay its high-redshift SF and related cosmic
events.  The No-Feedback model is also continuously affected by the
streaming velocity, while the other models show a disappearing $\vbc$
effect as reionization ends. 

\begin{figure}
\centering
\includegraphics[width=3.2in]{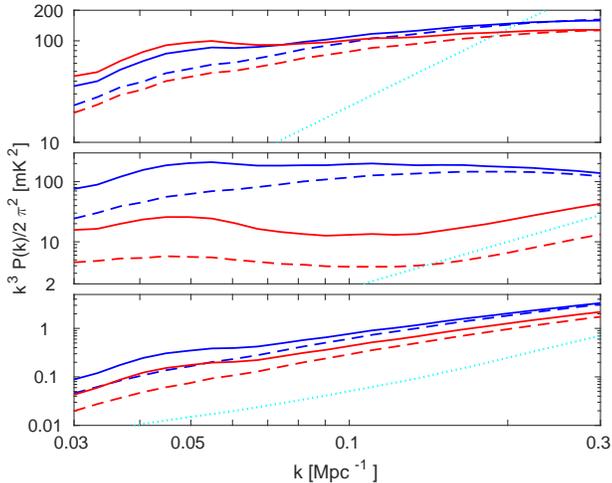}
\caption{{\bf The (spherically-averaged) 21-cm power spectrum as a
    function of $k$.} We consider different cases at several important
  redshifts, either with (solid) or without (dashed) the streaming
  velocity.  The top panel focuses on the Ly$\alpha$ peak, and
  compares the Maximal (red) and Gradual (blue) models. Corresponding
  redshifts are $z=27.4$ (red solid), 29.1 (red dashed), 24.7 (blue
  solid), and 25.5 (blue dashed). The middle panel focuses on the
  heating peak, and compares the Maximal (red) and Soft (blue) models.
  The redshifts are 19.2 (red solid), 19.6 (red dashed), 21 (blue
  solid), and 22 (blue dashed). The bottom panel shows the midpoint of
  reionization (i.e., when half the intergalactic medium has been
  reionized), and compares the Maximal (red) and No-Feedback (blue)
  models. The redshifts are 10.1 (red solid), 10.6 (red dashed), 8.4
  (blue solid), and 8.5 (blue dashed). The cyan dotted line (in each
  panel) shows the PS of the thermal noise for the SKA1 assuming a
  single beam, integration time of 1000 hours, and 10 MHz bandwidth at
  the redshift of the red solid line.}%%%
\label{fig:2}
\end{figure}

We show in Figure~\ref{fig:2} the PS versus $k$ at several key
redshifts. While the overall normalization (which is key for
observability) was shown in Figure~\ref{fig:1}, here we focus on the
PS shape (which is key for detecting the $\vbc$ signature).
Specifically, we compare the PS at $k=0.05~ \mpc^{-1}$ (the
large-scale BAO peak, where the $\vbc$ enhancement is maximized) to
its value at $k=0.3~ \mpc^{-1}$ (a scale that is small enough not to
be strongly affected by $\vbc$, yet is large enough compared to our
numerical resolution and the angular resolution of observations). We
thus examine (at each $z$) the ratio
\begin{equation}
  R \equiv \frac{\rm{PS}_{\vbc}(0.05)}{\rm{PS}_{\vbc}(0.3)}
  \left(\frac{\rm{PS}_{no\_\vbc}(0.05)}
    {\rm{PS}_{no\_\vbc}(0.3)}\right)^{-1}\ ,
\end{equation}
which compares the PS slope with and without $\vbc$. 

The top panel of Figure~\ref{fig:2} shows the Ly$\alpha$ peak, the
highest-redshift cosmic event with a large 21-cm signal \citep{BL05b}.
At this early time, low-mass halos are likely to dominate, so we
compare the Maximal model (with $f_*(M)$ giving a sharp cutoff)
%in the mass of star-forming halos 
to the Gradual model. As expected, the effect of
$\vbc$ is larger for the Maximal model ($R = 2.2$ compared to $R =
1.6$). Thus, early metal-cooling can enhance the BAO signature at the
Ly$\alpha$ peak.

The middle panel highlights the heating peak, showing the strong
dependence of this feature on the spectrum of the X-ray heating
sources. Hard photons travel further, 
%due to their lower cross section,
so the redshifting of the photon energies leads to slower cosmic
heating with hard X-rays, as well as smaller fluctuations due to the
large-scale smoothing of the heating \citep{Fialkov:2014}. The
stronger impact of fluctuations (including those from $\vbc$) in the
Soft case (as in 
% which is similar to the model of 
\citet{Visbal:2012})
yields $R = 2.9$, while for the Maximal (hard X-ray) case $R$ is only
1.4. We note that both cases in this panel assume the sharp $f_*(M)$
and no LW feedback, features made more likely by the possibility of
metal cooling.

The bottom panel shows the possibility of a strong BAO effect from
streaming velocity down to relatively low redshift.  Specifically, at
the midpoint of reionization (a time that current observations are
trying to probe), metal cooling may allow a large $\vbc$ effect. Since
photoheating feedback is significant at this time, we show the
No-Feedback case in addition to the Maximal model.  Ionization
fluctuations dominate here the 21-cm PS, and the photoheating feedback
makes the reionization more spatially uniform since regions with a
large ionized fraction experience stronger feedback\footnote{Note that
  at relatively early times the photoheating feedback can also enhance
  the fluctuations since it suppresses small galaxies leaving only
  larger, more highly biased ones.}; it also makes reionization more
gradual and thus pushes mid-reionization to a higher redshift relative
to the end of reionization. However, due to the continued SF in small
halos within the regions that have not yet been reionized, we find
only a small effect of this feedback on the power spectrum shape: $R =
1.8$ for the Maximal model compared with $R=2.0$ for the No-Feedback
case.  We note that in the case of H$_2$ cooling, LW feedback would
have saturated by this time (see section~\ref{Sec:Intro}), giving no
$\vbc$ effect (i.e., $R=1$), as for the Atomic case. 

For the Simulation-Based case, we get $R = 1.8$ at $z\sim14$
(ionization fraction of $\sim10\%$), while at lower redshifts small
halos are significantly suppressed and $R$ approaches 1. %%%

\section{Summary and Conclusions}
\label{Sec:sum}

In this Letter we have explored the possibility of metal-line cooling
in high-redshift low-mass halos, focusing on its effect on the
observable 21-cm signal from hydrogen. In the presence of metals,
small halos can cool and form stars even after the build-up of the LW
background (which shuts down SF via the H$_2$ cooling channel). Thus,
they can continue to contribute to SF even late in cosmic history.
Within this scenario we have made predictions for the 21-cm power
spectrum from a range of early times ($z = 6-40$), accounting for a
range of possible parameters for high-redshift astrophysical processes
that are poorly constrained at present. In particular, we considered
soft or hard SEDs of the first X-ray sources, a sharp or smooth cutoff
in the efficiency of SF in low-mass halos, and the effect of ionizing
photons on the amount of gas available for cooling (the photoheating
feedback).
%When modeling the photoheating feedback we generalized the
%prescription developed by \citet{Sobacchi:2013} to include partially
%ionized regions (and not only fully ionized ones), which resulted in
%stronger feedback.

We found that the SF in small halos, if it is indeed enhanced by metal
cooling, can strongly affect the 21-cm signal from all the considered
redshifts. An especially interesting implication is that the BAOs
imprinted by the supersonic relative velocities between dark matter
and baryons on the 21-cm signal, are enhanced and can survive to much
later times than previously thought, especially if SF in small halos
was more efficient than suggested by current simulations. In
particular, the height of the large-scale BAO peak (relative to the PS
at smaller scales) is enhanced for our
%%%
Simulation-Based model by a factor of $R = 1.8$ down to $z=14$ (the
beginning of reionization).  For the Maximal model the enhancement is
by a factor of $R = 2.2$ at the $z=27$ Ly$\alpha$ peak, $R=1.4$ at the
$z=19$ heating peak (boosted to $R = 2.9$ in the Soft X-ray case), and
a still significant $R = 1.8$ at the midpoint of reionization ($z=10$
in this model).

Current experiments are making rapid progress toward detecting the
21-cm signal from $z\sim 10$, while future experiments are planned for
even higher redshifts. These instruments will likely be able to probe
the role of small halos in the history of the early Universe by
measuring their 21-cm signature, in particular the BAOs imprinted by
the supersonic streaming velocity. Such an exciting detection would
shed light on the dawn of star formation.

\section{Acknowledgments}
We thank L. V. E. Koopmans for providing the noise spectrum of the
SKA. %%%
R.B.\ and A.C.\ acknowledge Israel Science Foundation grant 823/09 and
the Ministry of Science and Technology, Israel. A.F.\ was supported by
the LabEx ENS-ICFP: ANR-10-LABX-0010/ANR-10-IDEX- 0001-02 PSL. R.B.'s
work has been done within the Labex Institut Lagrange de Paris (ILP,
reference ANR-10-LABX-63) part of the Idex SUPER, and received
financial state aid managed by the Agence Nationale de la Recherche,
as part of the programme Investissements d'avenir under the reference
ANR-11-IDEX-0004-02. R.B. also acknowledges a Leverhulme Trust
Visiting Professorship.%%%

%\bsp

\end{document}